# Photoinduced charge density wave transition like a puppet on a string

Zhao-Jun Suo[1,2], Wen-Hao Liu[1,2], Zhi Wang[1], Hao-Wen Liu[1,2], Jun-Wei Luo[1,2,3]\*, Shu-Shen Li[1,2], and Lin-Wang Wang[4]\*

[1]*State Key Laboratory of Superlattices and Microstructures, Institute of Semiconductors, Chinese Academy of Sciences, Beijing 100083, China*
[2]*Center of Materials Science and Optoelectronics Engineering, University of Chinese Academy of Sciences, Beijing 100049, China*
[3]*Beijing Academy of Quantum Information Sciences, Beijing 100193, China*
[4]*Materials Science Division, Lawrence Berkeley National Laboratory, Berkeley, California 94720, USA*

\* jwluo@semi.ac.cn; lwwang@lbl.gov

## Abstract

Charge density wave (CDW) materials can undergo an ultrafast phase transition after an ultrashort laser pulse excitation, and the suggested underlying mechanisms have always been associated with two main features: excitonic interaction-induced CDW charge order and electron-phonon coupling-induced periodic lattice distortion (PLD). Here, beyond these two mechanisms, we reveal that photoexcitation induced CDW phase transition in the prototypic CDW example 1T-TiSe$_2$ is similar to a puppet on a string: six Ti-Se bonds connected to each distorted Ti atom acting as six strings controlling the PLD and in turn the CDW orders. The photoexcitation induced modulation on charge population of the Ti-Se bonds generates a laser-fluence-dependent interatomic-repulsive force along each Ti-Se bond. The nonequal length of these six Ti-Se bonds gives rise to a net force exerted on the central distorted Ti atom to push it toward the suppressing of the PLD and thus the CDW orders. We further illustrate that the dynamics of each distorted Ti atom behaves as though it attached to a spring in a simple harmonic motion with a fluence-dependent oscillation frequency, uniting two previously reported scaling laws for phase transition time. These findings significantly advance the understanding of CDW instability and provide new insights into how photoexcitation induced modulation on charge population may lead to phase transitions by directly connecting interatomic forces with reaction pathways.

**INTRODUCTION**

The complex interplay between the electronic, orbital, and lattice degrees of freedom in solids can give rise to a wide range of exotic phenomena, such as superconductivity [1,2] and charge density wave (CDW) states [3,4]. They are both cooperative electronic states associated with electron-phonon coupling and Fermi surface instabilities but exhibit vastly different macroscopic properties. Controlling and tuning many-body interactions in superconductivity and CDW, for example, through an electric field [5], can provide even more opportunities to manipulate the material properties. However, the complexity of the interplay often hinders us from revealing the essential degrees of freedom to gain deeper insight into the underlying microscopic physics. For instance, the origin of the CDW has been a subject of debate for decades [6–12]. Below the transition temperature $T_{CDW}$ = 202 K, the prototype CDW material 1T-TiSe$_2$ [13] possesses a CDW charge order accompanied by periodic lattice distortion (PLD) with the displacement of Ti atoms from their high-symmetry positions with a periodicity of (2×2×2) [14]. The current debate focuses on whether the CDW charge order is a consequence of the PLD structural order or the other way around [15]. The main argument is that the strong Jahn-Teller-like electron-phonon coupling creates a PLD, which modifies the electronic structure to give rise to the CDW charge order [6,7,16–18]. Another belief is that the strong Coulomb interaction between electrons and holes (or excitonic interactions) induces the condensation of electron-hole pairs and modulates the electron density to create the CDW charge order [19–21]. The formation of the CDW charge order in turn modifies the forces among the ions to generate a PLD [22]. Thus, the excitonic interactions alone are regarded as the driving force for the CDW transition [23–28], and it is hypothesized that ultrafast quenching of the CDW order is caused by increased dielectric screening under laser pulse irradiation [29].

A number of experiments have observed the robustness of PLD structural order against the quenching of CDW charge order triggered by photoexcitation of valence electrons [4,22]. In a recent simulation, the CDW charge order changed by only ~20% during the inversion of PLD structural order. These results suggest that the CDW transition is possibly a result of the cooperation of excitonic interactions and Jahn–Teller-like electron-phonon coupling [4,13,30–34]. Therefore, a consensus of opinion on the CDW transition has not yet been achieved. Specifically, these proposed mechanisms cannot explain the recently discovered partial recovery of the CDW order, which occurs at a longer time scale of approximately 1.5 ps following the initial quenching of the CDW charge order [35,36]. This implies that the driving force of the CWD transition may be strongly correlated with neither the CDW charge order nor the Jahn-Teller-like electron-phonon coupling. If this is true, all the abovementioned arguments will become invalid.

In this work, we provide a direct connection at atomic resolution between the structural and electronic dynamics responses to photoexcitation by performing rt-TDDFT simulations of the atomic dynamics of 1T-TiSe$_2$ under laser pulse irradiation. Our simulations excellently reproduce both the dynamics and timescales of the photoinduced CDW transition measured in experiments [33], enabling us to unravel the atomic force driving CDW instability and its partial recovery by identifying the related degrees of freedom. Contrary to previous arguments, we found that the driving force of the photoinduced CDW transition arises fully from the modulation of the charge population of six unequal-length Ti-Se bonds connected to each periodically distorted Ti atom, which exerts a net force on the distorted Ti atoms and pulls the to the high-symmetry position where the six Ti-Se bonds become equal in length and the net force vanishes. Since excitonic interactions and Jahn–Teller-like electron-phonon coupling are not the main driving forces, the CDW and PLD orders are fingerprints and not origins of the CDW phase. The photoinduced CDW transition behaves likes a puppet on a string with the six connected Ti-Se bonds acting as strings controlling the PLD and then CDW orders. We further illustrate that the partial recovery of CDW order at a longer time scale is due to the recombination of electron-hole pairs, which releases charge modulation-induced interatomic forces.

**RESULTS AND DISCUSSION**

**Ground state properties.** Bulk 1T-TiSe$_2$ at room temperature is in the semimetallic normal phase ($P\bar{3}m1$ [37]) with a hole pocket (valence band maximum, VBM) at the Γ point and electron pockets (conduction band minimum, CBM) at the L points of the Brillouin zone [29], as schematically shown in Fig. 1f. The Coulomb attraction between electrons and holes leads to the spontaneous formation of excitons, presumably followed by exciton condensation [38–40]. Figure 1a shows that each monolayer of 1T-TiSe$_2$ [11,41,42] consists of planes of hexagonally arranged Ti atoms sandwiched by two Se layers coordinating the central Ti atom in an octahedral arrangement. The six Ti-Se bonds in each octahedron have an identical length of 2.566 Å, and the Ti-Ti bonds have a bond length of $d_N = 3.538$ Å. Upon cooling below $T_{CDW}$ = 202 K [13], bulk 1T-TiSe$_2$ undergoes a transition into a commensurate CDW state accompanied by a PLD with the displacement of Ti atoms from their high-symmetry positions with a periodicity of (2×2×2) ($P\bar{3}c1$ [37]). The reduction of the Brillouin zone in the CDW phase folds the CBM to the zone center and opens a bandgap at the Fermi level due to anticrossing between the downward-dispersing conduction band and upward-dispersing valence band [10,29]. The PLD results in two types of Ti-centered octahedra [42], as shown in Fig. 1c-e. In one type, as shown in Fig. 1e, the central Ti atom (named Ti(II)) remains intact, but the top 3 Se atoms displace in a clockwise rotation, and the bottom 3 Se atoms displace in a counterclockwise rotation, leaving the bond lengths of all six Ti-Se bonds unchanged. In another type, as shown in Fig. 1d, the central Ti atom (named Ti(I)) has an off-center shift, breaking the equal bond lengths of the six Ti-Se bonds into three groups. The off-center shift of the Ti(I) atom also alters the Ti-Ti bond length of $d_N = 3.538$ Å in the normal phase into short Ti-Ti bonds with a bond length of $d_S = 3.451$ Å and long Ti-Ti bonds with a bond length of $d_L = 3.625$ Å. Thus, the bond lengths of Ti-Ti bonds characterize the structural order of the CDW phase.

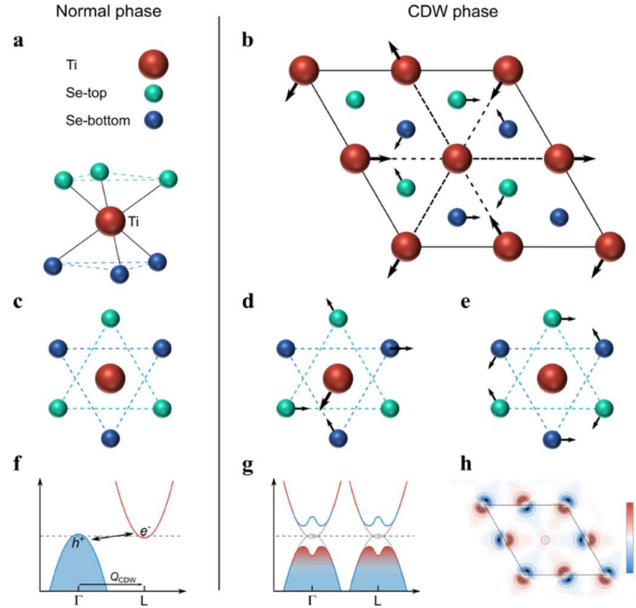

**FIG.1. Atomic configurations of monolayer 1T-TiSe$_2$ in the normal and CDW phases. a,**

TiSe$_6$ octahedra in the normal phase. **b**, Primitive cell of the CDW phase with PLD ($2 \times 2$ superlattice). The arrows stand for the displacements. **c**, TiSe$_6$ octahedron in the normal phase. **d**, **e**, Ti(I)Se$_6$ and Ti(II)Se$_6$ octahedra in the CDW phase. **f**, Electronic structure of the normal phase. **g**, Electronic structure of the CDW phase. **h**, Calculated CDW pattern of the charge density obtained according to the formulas given in Ref. [31].

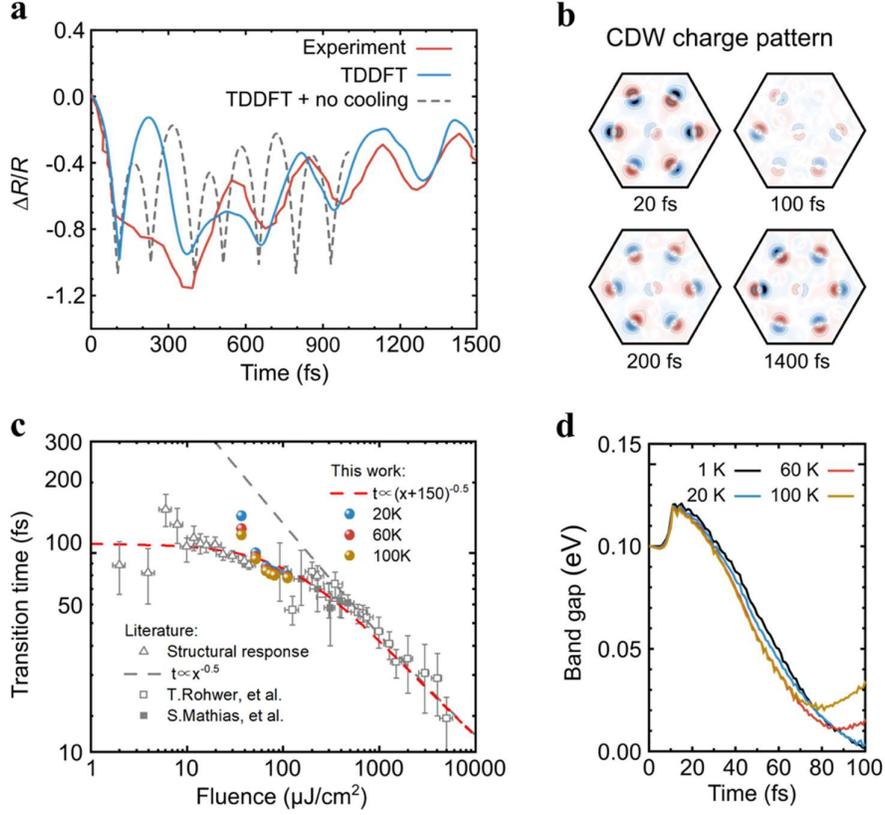

**FIG.2. Photoexcitation-induced ultrafast CDW transition. a**, Time-dependent transient reflectivity change $\Delta R/R$ from experiment (at 10 K) [33] and simulation (at 20 K). In order to comparison, here we also include the simulation result (dashed grey line) using the conventional TDDFT without considering hot carrier cooling process. **b**, The simulated CDW charge order at different times following the photoexcitation. **c**, Simulated transition time $\tau_{CDW}$ (colored dots) as a function of laser fluence under different initial temperatures of 20, 60, and 100 K, respectively, in comparison with that obtained from three experimental measurements. Two different scaling laws have been suggested for transition times $\tau_{CDW}$: the experimental data exhibit a scaling law $\tau_{CDW} \propto 1/\sqrt{n_{photon}}$ for laser fluence above 100 μJ/cm$^2$ [33] and follows $\tau_{CDW} \propto 1/n_{photon}$ if laser fluences below 200 μJ/cm$^2$ [43]. Here, we can unit them into a single formula $\tau_{CDW} \propto 1/\sqrt{n_{photon} - n_{th}}$ (red dashed line) based on a harmonic oscillation model (details see main text). **d**, Simulated CDW gap quenching after photoexcitation at different temperatures.

**Photoexcitation-induced phase transition.** Figure 2 shows the temporal dynamics of the photoinduced phase transition obtained from rt-TDDFT simulation for the CDW phase 1T-TiSe$_2$ under irradiation of a 20-fs laser pulse at an initial temperature of 20 K. We tuned the electric field $E_0$ of the mimicked laser pulse (see "Methods summary") to excite 0.9% of the valence electrons (3.6 electrons in a supercell) from the valence bands to the conduction bands, corresponding to a fluence of approximately 52 µJ/cm$^2$, with the same magnitude as that adopted in the experiment [43]. To monitor the lattice dynamics, Figure 2a shows the calculated transient reflectivity change $\Delta R/R$ [44] of the lattice following photoexcitation according to Eq. (2) with atom positions obtained from the rt-TDDFT simulation (see "Methods summary"). The calculations are in reasonable agreement with the experimental data [33] over a wide time window spanning from 0 to 1.5 ps. The discrepancies that appear in a small segment spanning from 100 to 300 fs may be attributed to the failure to experimentally probe the signal responsible for the transition from the normal phase to the inverted PLD since a recent experiment [30] gives better agreement with our simulation of this curve shape. This distinct will be discussed in detail later. Following photoexcitation, $\Delta R/R$ exhibits a drop corresponding to the suppression of the PLD. Such a drop is concomitant with a fast reduction in the CDW gap, as shown in Fig. 2d. The first dip in the $\Delta R/R$ curve occurs at approximately 100 fs, indicating the completion of the phase transition with a fully suppressed PLD structural order in the normal phase. The corresponding time of the first dip is called the drop time in some experimental literature [23,29,43] and, is here named the transition time. Fig. 2b shows that at a transition time of 100 fs, the CDW charge order is almost fully quenched along with complete closure of the CDW gap (Fig. 2d). Therefore, we have shown that both the CDW charge order is quenched and the CDW gap is closed at approximately 100 fs along with full suppression of the PLD.

After the completion of the phase transition at 100 fs, the rt-TDDFT simulation-predicted $\Delta R/R$ bounces back and reaches a peak at 200 fs (Fig. 2a). The corresponding CDW charge pattern shown in Fig. 2b is an inversion of the original CDW pattern at 20 fs. This illustrates that the peak at 200 fs is responsible for the inverted PLD. The times for the transition from the CDW phase to the normal phase and from the normal phase to the CDW phase with an inverted PLD are both 100 fs.

The oscillation of the $\Delta R/R$ curve corresponds to a lattice oscillation with a frequency of 3.5 THz, which is in good agreement with the experimentally observed $A_{1g}^*$ CDW amplitude mode with a frequency of 3.4 THz that emerged during the photoexcitation-induced CDW phase transition [4,36,45,46]. A recent experiment [33] revealed the recovery of the CDW order at a long time scale after a photoinduced phase transition, evidenced by damped oscillations with an increase in the reflectivity change $\Delta R/R$ from 300 fs to 1500 fs, as shown in Fig. 2a. Our rt-TDDFT simulation also reproduces such long-time-scale experimental results well.

Photoinduced phase transitions have been reported to be very sensitive to laser fluence: higher laser fluences usually give rise to shorter transition times [23,29]. In Fig. 2c, we show the transition time as a function of laser fluence obtained from our rt-TDDFT simulations as well as experimental data reported by three different groups [23,29,43] for a variety of laser fluences and temperatures. Our predicted transition times are in excellent agreement with the experimental results over a wide range of laser fluences and temperatures, demonstrating the robustness of our rt-TDDFT simulations against changes in the laser pulse and temperature. Consequently, we simultaneously reproduced the photoexcitation-induced CDW phase transition in both structural and electronic orders by rt-TDDFT simulations. This provides a solid foundation for our following analysis of the transition mechanism.

**Atomic disorder during phase transition.** The transient reflectivity change $\Delta R/R$ (shown in Fig. 2a) and X-ray diffraction measurements reflect the averages over many unit cells (see Eq. (2) in the "Methods summary" section) and are thus less sensitive to random displacements of individual atoms [47]. Using total scattering of femtosecond X-ray pulses, Wall et al., [47] recently observed atomic disordering during a photoinduced phase transition in vanadium dioxide ($VO_2$) and concluded that atomic disordering is central to the transition mechanism. To directly evaluate the effect of atomic disorder, here, we examine the photoinduced atomic dynamics under various temperatures by considering thermal excitations of different phonon modes. Figure 3 shows the rt-TDDFT evolution of all Ti-Ti bonds after photoexcitation for initial temperatures of 1, 20, 60, and 100 K. We indeed observe that, except for the 1 K case, the structural transition proceeds with the uncorrelated disordering of Ti ions from their initial PLD distribution rather than via previously proposed cooperative motion in a collective optical phonon mode [48–50]. In the case of 1 K, Fig. 3a shows that after photoexcitation, long Ti-Ti bonds exhibit a reduction in their bond lengths, whereas short Ti-Ti bonds exhibit elongation, converging quickly toward an identical bond length ($d_N$ = 3.538 Å), as expected for the transition to the normal phase. All bonds cross at $t \sim 100$ fs (reaching the normal phase) and overshoot the normal phase bond length before rebounding at $t \sim 150$ fs, resulting in a substantially damped oscillation. However, Fig. 3b shows that as the temperature rises to 20 K, the photoexcited dynamics of Ti-Ti bonds have strong uncorrelated disorder without a common phase and frequency, which is different from the coherent collective behavior in the 1 K case (Fig. 3a). This uncorrelated disorder of Ti ions is remarkably intensified by a further increase in temperature, as shown in Fig. 3c and 3d for the cases of 60 and 100 K, respectively. The temperature-dependent atomic disorder implies that thermally excited intrinsic phonon modes may play important roles in photoexcited atomic dynamics.

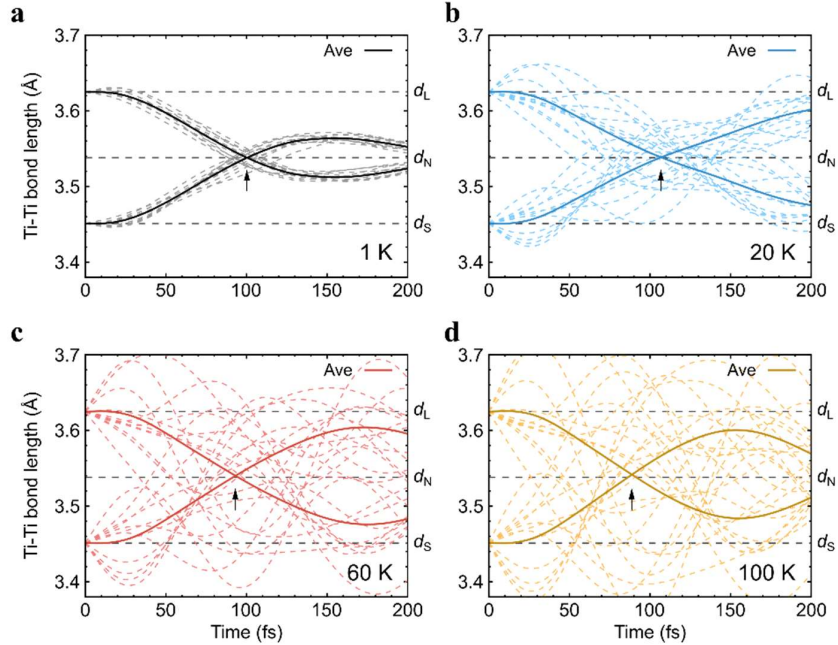

**FIG.3. Temperature dependence of photoexcited atomic dynamics from rt-TDDFT simulations.** Length evolution of short Ti-Ti bonds ($d_S$) and long Ti-Ti bonds ($d_L$) at initial temperatures of **a**, 1 K, **b**, 20 K, **c**, 60 K, and **d**, 100 K using an identical laser pulse. The dashed lines in the background indicate the lengths of all individual Ti-Ti bonds in the supercell, and the solid lines represent the means of $d_S$ and $d_L$, which are averages over all short Ti-Ti bonds and long Ti-Ti bonds according to Fig. 1b, respectively. The vertical arrows indicate the completion of the phase transition.

Such atomic disorder is, however, expected to be absent in the transient reflectivity change $\Delta R/R$ and X-ray diffraction results since these values reflect the averages over many unit cells and are insensitive to random displacements [47]. To exclude atomic disorder, we therefore examine the dynamics of the average bond length over long Ti-Ti bonds and short Ti-Ti bonds, as shown in Fig. 3. Indeed, the dynamic evolution of average bond lengths exhibits nearly the same tendency among the four different temperatures, although the movement of individual ions is rather sensitive to temperature. Remarkably, all cases show transition times at $t \sim 100$ fs. The discrepancy occurs mainly after the completion of the phase transition, manifesting mainly as a difference in the damping of the oscillation. Surprisingly, the oscillation of the average bond length for higher temperatures shows smaller damping than in the 1 K case, indicating that higher temperature might make CDW inversion easier. In the 20 K case, at 200 fs before rebounding, the average bond length of long (short) bonds reaches the short (long) bond length in the original CDW phase within only an ~0.02 Å difference, thus realizing almost perfect PLD inversion. This inverted PLD is responsible for the second peak (at $t \sim 200$ fs) in the simulated $\Delta R/R$, as shown in Fig. 2a. However, such a peak is absent in the experimental measurements where a line connects two dips. This is slightly abnormal because a peak of the PLD order is expected between the two dips as the dips correspond to the normal phase. Therefore, we attribute this partial discrepancy between the experimental result and rt-TDDFT simulation to the possible failure of experimentally probing the signal responsible for the PLD inversion. Interestingly, our simulation-predicted second peak was confirmed in a very recent experiment [30].

We have demonstrated that the dynamic evolution of the average bonds is insensitive to temperature-dependent atomic disorder, ruling out atomic disorder as a central factor in the phase transition mechanism [47]. Moreover, the photoinduced CDW transition is usually suggested to be driven by electron-phonon interactions [6,16,20,31,51]. This assumption is justified by the experimental observation [4] that a higher temperature is required to quench the PLD even if CDW excitonic correlations have been melted by the laser. In the above discussion, we have shown that thermally excited lattice vibrations exist in photoinduced atomic dynamics. However, after the thermal vibrations have been averaged out, the pure phase transition dynamics become relatively temperature independent. We can thus assume that laser-induced electron excitation induces a CDW amplitude mode [45,52] that has no strong coupling with thermally excited lattice vibrations. This rules out intrinsic phonons and associated phonon-phonon scattering (through heating) as the main cause for the photoinduced CDW phase transition.

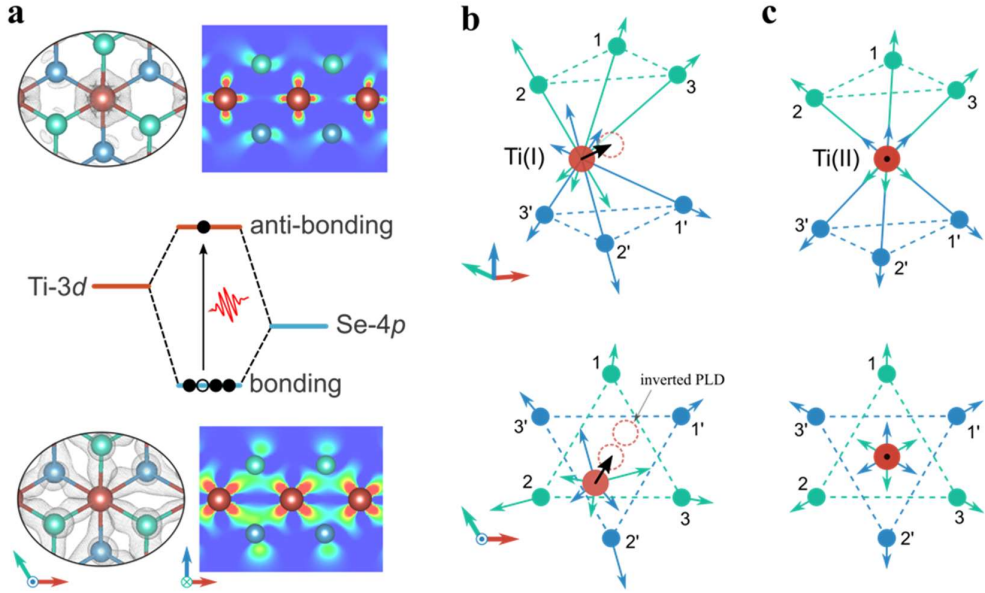

**FIG.4. Photoexcited free charge carriers and the resulting atomic forces. a**, Charge density of photoexcited free electrons (top panels) and holes (bottom panels) obtained by plotting 3D isosurfaces and 2D maps. The 2D plane is chosen to contain one Ti(I) atom and its four bonded Se atoms (two long Ti-Se bonds and two short Ti-Se bonds) to examine the charge distribution of the Ti-Se bonds. The middle panel is a schematic diagram of the photoexcitation of valence electrons from the bonding states (valence band), leaving behind positively charged holes, to the antibonding states (conduction band) of Ti $3d$ and Se $4p$ orbitals. **b**, Schematic diagram of the distorted Ti(I)Se$_6$ octahedron with photogenerated interatomic forces along each Ti-Se bond arising from modified charge population. Blue arrows indicate the interatomic forces of Ti(I) bonded to Se atoms in the bottom layer, and green arrows indicate the forces of Ti(I) bonded to Se atoms in the top layer. The black arrow indicates the net force exerted on the Ti(I) atom. This atomic force pulls the Ti(I) atom, causing it to overshoot the position in the normal phase and even reach the position in the inverted PLD. **c**, Same as in **b** but for the distorted Ti(II)Se$_6$ octahedron, where the interatomic forces of six Ti(II)-Se bonds are the same in magnitude and result in a zero net force exerted on the Ti(II) atom, as indicated by a dot. In (b) and (c), numbers 1, 2, and 3 mark the three Se atoms of the top layer, ordered in the counterclockwise direction, and 1', 2', and 3' mark the three Se atoms of the bottom layer, ordered in the clockwise direction.

**Photoexcited atomic forces driving the phase transition.** We now turn to unravel the atomic forces driving the CDW transition upon application of photoexcitation. It is generally believed that the excitonic interaction or electron-hole Coulomb attraction alone drives the CDW transition [9,23–27] considering that the CDW charge order is a result of the condensation of electron-hole pairs [32]. However, this many-body force alone cannot drive the transition, as illustrated by the experimental observation of 1T-TiSe$_2$ in which PLD structural order can persist even when the CDW charge order is quenched under irradiation with a laser pulse [4]. Although both the CDW charge order and PLD structural order are characteristics (fingerprints) of the CDW phase, do we have to associate the driving force with these two orders? One can perhaps view the problem from a different point of view in single-particle state of charge occupation. To the best of our knowledge, the relationship between the single-particle state occupation and atomic driving force for the CDW phase transition in 1T-TiSe$_2$ has not yet been shown. In 1T-TiSe$_2$, the top valence bands and bottom conduction bands are composed mainly of the bonding and antibonding states of Ti 3$d$ and Se 4$p$ orbitals, respectively [7]. Laser pulses excite electrons from the filled bonding states in the valence bands (leaving behind holes) to populate the empty antibonding states in the conduction bands. Figure 4a shows the partial charge density of photoexcited electrons and holes. One can see from their spatial distributions that both photoexcited electrons and holes are located mainly around the Ti atoms in a $d$-like shape. In addition, the photoexcited holes have a finite component on the Ti-Se bonds, manifesting a bonding character, whereas the photoexcited electrons have zero components on the Ti-Se bonds, manifesting an anti-bonding character. When excited, the electron occupation of the previously empty Ti-Se antibonding band and hole occupation of the previously filled Ti-Se bonding band will gain energy by elongating the Ti-Se bond, which lowers the energy levels of the antibonding states and raises the energy levels of the bonding states, as schematically shown in Fig. 4a. Thus, photoexcitation effectively produces a pair of interatomic forces to stretch each Ti-Se bond out. The magnitude of this interatomic stretching force has a scaling rule of $F_{\text{Ti-Se}} \propto 2n_{\text{photon}}/d_{\text{Ti-Se}}^4$ [53], where $n_{\text{photon}}$ is the number of photoexcited electrons on an individual Ti-Se bond with length $d_{\text{Ti-Se}}$ and the prefactor 2 accounts for photoexcited holes.

Because six Ti(II)-Se bonds in each Ti(II)Se$_6$ octahedron have an identical bond length $d_{\text{Ti-Se}}$, six pairs of photoexcitation-generated interatomic stretching forces along the Ti-Se bonds are equal in magnitude and are paired into three groups (Ti(II)-Se1 versus Ti(II)-Se2', Ti(II)-Se2 versus Ti(II)-Se1', and Ti(II)-Se3 versus Ti(II)-Se3'), as shown in Fig. 4c. The two forces within each group on the Ti(II) atom are antiparallel with the same magnitude. Thus, the net force on the center Ti(II) atom is zero, resulting in the Ti(II) atoms remaining fixed during the CDW transition. However, in each Ti(I)Se$_6$ octahedron, the shift of the center Ti(I) away from the ideal position not only breaks the antiparallelism of the two forces of each group but also makes their force magnitude unequal. The PLD modulates the bond lengths of these six Ti(I)-Se bonds into three equal-length pairs: Ti(I)-Se1 and Ti(I)-Se1', Ti(I)-Se2 and Ti(I)-Se2', and Ti(I)-Se3 and Ti(I)-Se3', which have bond lengths of 2.651 Å, 2.493 Å, and 2.564 Å, respectively. Therefore, the two forces for each pair have the same magnitude but are not exactly antiparallel, as schematically shown in Fig. 4b. As a result, the two forces give rise to a finite net force on the Ti(I) atom. The three in-plane net forces of the three pairs result in a net force with a direction antiparallel to the direction of the PLD. This total net force pushes each Ti(I) atom toward its high-symmetry position in the normal phase, in which the net force tends to vanish, as all six Ti(I)-Se bonds have an identical bond length, the same as the case in the Ti(II)Se$_6$ octahedron. At this high-symmetry position (the equilibrium position), the acceleration of the Ti(I) atom falls to zero, but the velocity is at its maximum, resulting in the Ti(I) atom moving farther in the same direction. This explains why all Ti-Ti bonds overshoot the normal phase before rebounding after the transition time t = 100 fs, as shown in Fig. 3. The overshoot makes Ti(I)-Se1 and Ti(I)-Se1' bonds shorter than Ti(I)-Se2 and Ti(I)-Se2' bonds, creating a restoring force on Ti(I) atoms (see Fig. 4b). The restoring force increases as the overshooting displacement increases. The Ti(I) atom has zero velocity at the maximum overshooting displacement, which, as shown in Fig. 3, is sensitive to lattice temperature. At high temperatures, the maximum overshooting displacement of Ti(I) atoms inverts the PLD structural order, as indicated in the bottom panel of Fig. 4b. We thus demonstrate that each Ti(I) atom behaves as though attached to a spring in a simple harmonic manner. It is thus the momentum of these Ti(I) atoms that pushes the system into a dynamic oscillation mode, hence causing the inversion of the PLD and CDW order. If there is no damping (i.e., photoexcited carriers remain unchanged without losing their energy to the lattice via phonons), this oscillation will continue. Such coherent oscillations of the Ti(I) atoms generate the experimentally observed in-plane

CDW phonon mode with a frequency of approximately 3.4 THz [4,36,45,46]. Therefore, we have unambiguously illustrated that the photoinduced CDW phase transition acts like a puppet on a string with six Ti(I)-Se bonds acting as strings to modulate the PLD structural order without invoking excitonic interactions and Jahn–Teller-like electron-phonon coupling. Both the CDW charge order and PLD structural order are fingerprints rather than the origins of the CDW phase.

The above explanation implies that the temperature can also play a role in the laser-induced CDW phase transition since there are also thermally excited charge populations in the Ti-Se bonds. Figure 2c shows that the transition time is sensitive to temperature at weak laser fluences but becomes insensitive with increasing fluence. This phenomenon can be understood based on the relative ratio between thermally excited and photoexcited charge populations. The thermally excited charge population is proportional to temperature, whereas the photoexcited population can be enhanced by laser fluence. At weak laser fluences, the thermally excited charge population is appreciable in comparison with the photoexcited population. However, at strong laser fluences, the thermally excited charge population becomes negligible, and thus, the transition time becomes insensitive to changes in temperature.

**Fluence-dependent transition time.** We can now explain why a higher fluence laser pulse has a shorter CDW transition time, as shown in Fig. 2c. Our rt-TDDFT simulations predict $\tau_{\text{CDW}} \propto 1/n_{\text{photon}}$ in good agreement with recent experiments [43] in which the laser fluence is < 0.2 mJ/cm$^2$. However, one early experiment using high fluence laser pulses (> 0.1 mJ/cm$^2$) showed $\tau_{\text{CDW}} \propto 1/\sqrt{n_{\text{photon}}}$. We have demonstrated above that, in the coherent oscillations of the Ti(I) atoms, each Ti(I) atom is a simple harmonic oscillator with an equilibrium position in the normal phase, and the maximum displacement is at the PLD or inverted PLD (if damping is neglected). A simple harmonic oscillator will oscillate with equal displacement on either side of the equilibrium position if the net force can be described by Hooke's law $F = kx$ ($k$ is the force constant and $x$ is the displacement). Here, the equilibrium position for the Ti(I) atom is its position in the normal phase. The frequency of a simple harmonic oscillator is $\omega = \sqrt{k/m}$, where $m$ is the mass of the Ti atom. As discussed above, the maximum net force on the Ti(I) atom is $F_{\text{max}} \propto n_{\text{photon}} - n_{th}$ (where $n_{th}$ is the photoexcited electron density threshold for transition); thus, the force constant k is $k = F_{\text{max}}/\delta \propto (n_{\text{photon}} - n_{th})/\delta$, where δ is the displacement from the normal phase to the CDW phase. We thus obtain $\omega_{CDW} \propto \sqrt{n_{photon} - n_{th}}$. The CDW transition time is the motion during one-quarter of the CDW phonon mode: $\tau_{CDW} \propto 1/\sqrt{n_{photon} - n_{th}}$. In Fig. 2, we show that we can unit different experimental data with two distinct scaling laws of $\tau_{CDW} \propto 1/\sqrt{n_{photon}}$ (for strong laser fluence) [29] and $\tau_{CDW} \propto 1/n_{photon}$ (for weak laser fluence) [43] into a single formula $\tau_{CDW} \propto 1/\sqrt{n_{photon} - n_{th}}$.

In Fig. 2c, we also note that in the weak laser fluence range, our simulations predicted a temperature-dependent $\tau_{CDW}$: a shorter $\tau_{CDW}$ at a higher temperature. However, in the strong laser fluence range, our simulations show that $\tau_{CDW}$ is temperature independent. We infer that this difference arises from the thermal population of free electrons and holes, which is expected to be in the same order as that of photoexcitation in the weak laser fluence range. As the laser fluence increases, the ratio of the thermal population decreases and finally becomes negligible under a strong laser fluence. This understanding should shed new light on the thermal-induced CDW phase transition.

**Physical mechanism underlying PLD recovery.** Photoexcited hot carriers usually cool down toward the band edge after a hundred femtoseconds through the emission of phonons in solids. If the size of the bandgap is smaller than the phonon energies, the photoexcited electrons will nonradiatively recombine with the photoexcited holes crossing the Fermi energy [54]. Figure 5 shows that such a nonradiative recombination process causes a reduction in $n_{photon}$, weakening the six stretching forces exerted on the Ti(I) atom. We thus expect the oscillation of Ti(I) to decay (be damped) due to this process. Indeed, experiments do reveal a partial recovery of PLD, as shown in Fig. 2a. However, the two-temperature model following Ref. [55] has been used to suggest that this recovery is mainly driven by thermalization of the electronic subsystem with the atomic lattice [33].

To gain insight into the hot carrier cooling and nonradiative recombination effects, in Fig. 2a, we also show the results of the conventional rt-TDDFT simulation without including the carrier cooling process. The simulation exhibits lattice oscillation with no damping and thus fails to repeat the experimental results. In the conventional rt-TDDFT simulation, the photoexcited electrons stay in their high-energy excited states all the time. Thus, there is no decay in photogenerated interatomic forces. The newly developed rt-TDDFT introduces the hot carrier cooling process by using a Boltzmann factor in the dynamics calculation [55]. Using this Boltzmann rt-TDDFT simulation, the photoexcited hot electrons and holes start to decay into lower energy states after 100 fs by emitting phonons, which raise the lattice temperature. Figure 5 shows that the hot electrons relax much faster than the hot holes, and the cooling rate slows at the end of our simulation because the carriers gather at the band edges due to the recovery of the CDW gap. However, we do find that the electrons at band-edge states will pass across the Fermi level to recombine nonradiatively with holes. Electron-hole recombination causes a reduction in the photogenerated interatomic forces. We expect that all photodoped free electrons and holes will ultimately disappear by the nonradiative recombination process on a longer timescale beyond that of our simulation. This time scale observed in experiments is much longer than 1.5 ps [33,35,36].

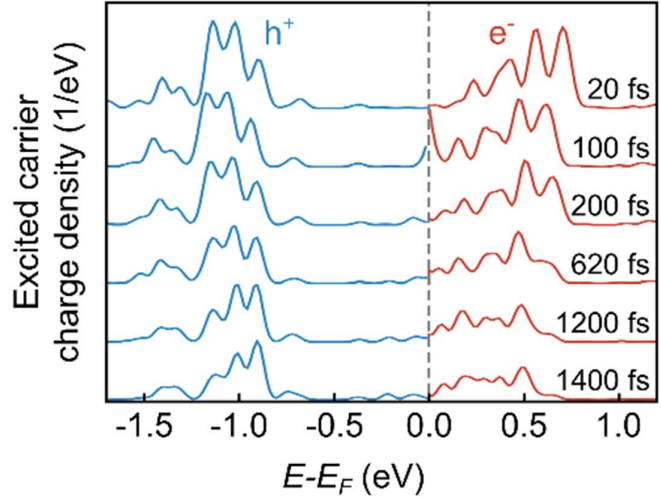

**FIG.5. Cooling of hot carriers following photoexcitation.** Snapshots of the partial density of states occupied by photoexcited electrons and holes for times of 20, 100, 200, 620, 1200, and 1400 fs following photoexcitation for the case shown in Fig. 2a. The vertical dashed line represents the Fermi level.

**CONCLUSION**

We simulated photoinduced CDW quenching (~100 fs) and recovery in monolayer 1T-TiSe$_2$, and the results agreed well with experimental results [33,43]. Moreover, we used the atomic driving force caused by the occupation of the antibonding state and unoccupation of the bonding state to explain the phase transition from a single-particle viewpoint. This explanation is different from previously proposed mechanisms in the literature [4,31]. The redistribution of photogenerated carriers exerts an interatomic force along Ti-Se bonds with a magnitude proportional to $\sim 1/d^4$ and affects the distortion in TiSe$_6$ octahedra, which regulates the dynamics of Ti-Ti bonds with a frequency of 3.5 THz in accordance with the experimentally observed 3.4 THz $A_{1g}^*$ CDW amplitude mode. The atomic driving force depends on the fluence (photogenerated carrier number) and the bond length. This dependence causes the CDW quenching time to have the following relation: $\tau_{\text{CDW}} \propto 1/\sqrt{n_{\text{photon}} - n_{th}}$, which units two previously suggested scaling laws of $\tau_{\text{CDW}} \propto 1/\sqrt{n_{\text{photon}}}$ (for strong laser fluence) [29] and $\tau_{\text{CDW}} \propto 1/n_{\text{photon}}$ (for weak laser fluence) [43]. During the carrier cooling and nonradiative recombination process, the CDW phase gradually recovers due to the reduction of excited carriers and the corresponding decrease in the atomic driving force.

**Methods summary**

In the rt-TDDFT simulations of the photoinduced CDW transition in 1T-TiSe$_2$, a 48-atom supercell of monolayer 1T-TiSe$_2$ with a 12 Å vacuum is used to prevent interaction with periodic images. We adopt the optimized norm-conserving Vanderbilt pseudopotentials (ONCV) from SG15 [56] with a plane wave kinetic energy cutoff of 65 Ry within the Perdew-Burke-Ernzerhof (PBE) generalized gradient approximation (GGA) for the exchange-correlation functional. A Γ-centered $4 \times 4 \times 1$ k-mesh is used based on the convergence tests. Our DFT calculation predicts that at zero temperature, the lattice constant of the CDW phase is 3.538 Å, and the PLD-caused atomic displacements of Ti and Se atoms are 0.087 Å and 0.029 Å, respectively, in line with those reported in the literature [13,42]. The predicted bandgaps of monolayer 1T-TiSe$_2$ in the normal and CDW phases are -0.430 eV and 0.087 eV, respectively, very close to those reported in Ref. [42] (-0.446 eV and 0.082 eV).

We utilize a recently developed rt-TDDFT algorithm by promoting the ion step $\Delta t$ from the usual 0.001 fs to 0.1~0.5 fs [57] as implemented in the PWMAT package [32] to capture the atomic dynamics following photoexcitation. In our rt-TDDFT simulations, the system is under irradiation with an electronic field $E(t) = E_0 \sin(\omega t) \exp[-\frac{(t-t_0)^2}{\sigma^2}]$ (where $\omega = 0.7\,\pi/\text{fs}, t_0 = 10$ fs, and $\sigma = 4$ fs), which corresponds to a 20-fs duration laser pulse with a photon energy of 1.45 eV, consistent with experimentally utilized 20-fs laser pulses with photon energies of 1.57 eV [29] and 1.55 eV [23]. For direct comparison with experimental measurements of structural order, we calculate the transient reflectivity change $\Delta R/R$ [44] according to

$$\frac{\Delta R}{R} \approx A \sum_{i \in \text{Ti}} [x_i^2(t) - x_{i,0}^2], \qquad (2)$$

where $x_i(t)$ and $x_{i,0}$ are the time-dependent coordinates and initial position of each Ti atom, respectively, and $A$ is a normalization factor to keep the $\Delta R/R$ of the normal phase equal to 1. We take the atom coordinates in the CDW phase as the reference coordinates. In the conventional rt-TDDFT based on Ehrenfest dynamics [57,59], the atoms are regarded as classical particles described by Newton's theories, while the electron wavefunctions are treated quantum mechanically. The Ehrenfest-TDDFT fails to treat the hot carrier cooling process since it lacks a detailed balance [55]. To correct this, we introduce a Boltzmann factor to the conventional Ehrenfest rt-TDDFT [55] by adding a change $\Delta C_{i,l}(t)$ to the wavefunction coefficient. We have used a dephasing factor of $\tau_{i,j} = 20$ fs in the Boltzmann correction. For a more detailed description of the Boltzmann factor formalism, refer to Ref. [55].

## ACKNOWLEDGMENTS

The work in China was supported by the Key Research Program of Frontier Sciences，CAS under Grant No. ZDBS-LY-JSC019, CAS Project for Young Scientists in Basic Research under Grant No. YSBR-026, the Strategic Priority Research Program of the Chinese Academy of Sciences under Grant No. XDB43020000, and the National Natural Science Foundation of China (NSFC) under Grant Nos. 11925407 and 61927901. L.W. W was supported by the Director, Office of Science, the Office of Basic Energy Sciences (BES), Materials Sciences and Engineering (MSE) Division of the U.S. Department of Energy (DOE) through the theory of material (KC2301) program under Contract No. DEAC02-05CH11231.